\documentclass[letterpaper, 10 pt, conference]{IEEEtran}  

\IEEEoverridecommandlockouts                              

\usepackage[utf8]{inputenc}
\usepackage[T1]{fontenc}
\usepackage{cite}
\usepackage[dvips]{color}
\usepackage{tabu}
\usepackage{comment}
\usepackage{todonotes}
\usepackage{epsf}
\usepackage{epsfig}
\usepackage{times}
\usepackage{epsfig}
\usepackage{graphicx}

\usepackage{mathtools}
\usepackage{mathrsfs}
\usepackage{amssymb,amsmath}
\usepackage{amsmath}
\usepackage{amsfonts}

\usepackage{cite}

\usepackage{amsmath,amssymb,amsfonts}
\usepackage{graphicx}
\usepackage{textcomp}
\usepackage{xcolor}
\usepackage{soul}
\usepackage{arydshln}
\usepackage{makecell}
\usepackage{svg}

\usepackage[algo2e]{algorithm2e} 
\usepackage{cleveref}
\usepackage{dsfont}
\usepackage{lettrine} 
\usepackage{epsfig,algorithm,algpseudocode,amsthm,url}
\usepackage{caption}
\usepackage{subcaption}
\usepackage[justification=raggedright]{caption}
\usepackage{subcaption}
\usepackage{bbm}
\usepackage[utf8]{inputenc}
\usepackage{verbatim}
\usepackage{setspace}	
\usepackage[T1]{fontenc}
\usepackage{textcomp}
\usepackage{tabularx}
\usepackage{xpatch}

\usepackage{stackengine}
\usepackage[nocomma]{optidef}
\usepackage{commath}

\title{\LARGE \bf
Deciphering Heartbeat Signatures: A Vision Transformer Approach to Explainable Atrial Fibrillation Detection from ECG Signals
}

\author{Aruna Mohan$^{1}$, Danne Elbers$^{1,2}$, Or Zilbershot$^{1}$, Fatemeh Afghah$^{3}$, David Vorchheimer$^{4}$
\thanks{$^{1}$A. Mohan, D. Elbers, and O. Zilbershot are with Walkky,
        Boston, MA, USA.
        {\tt\small \{aruna,danne,or\}@walkky.com}}%
\thanks{$^{2}$D. Elbers is with the Chobanian \& Avedisian School of Medicine, Boston University, Boston, MA 02118, USA.}
\thanks{$^{3}$F. Afghah is with the Department of Electrical \& Computer Engineering, Clemson University,
        Clemson, SC 29631, USA.
        {\tt\small fafghah@clemson.edu}}%
\thanks{$^{4}$D. Vorchheimer is with Northwell Health Physician Partners, Cardiology, 158 East 84th Street, New York, NY 10028, and Lenox Hill Hospital, 100 E 77th St, New York, NY 10075, USA.
        {\tt\small dvorch@gmail.com}}%
}

\begin{document}

\maketitle
\thispagestyle{empty}
\pagestyle{empty}

\begin{abstract}

Remote patient monitoring based on wearable single-lead electrocardiogram (ECG) devices has significant potential for enabling the early detection of heart disease, especially in combination with artificial intelligence (AI) approaches for automated heart disease detection. There have been prior studies applying AI approaches based on deep learning for heart disease detection. However, these models are yet to be widely accepted as a reliable aid for clinical diagnostics, in part due to the current black-box perception surrounding many AI algorithms. In particular, there is a need to identify the key features of the ECG signal that contribute toward making an accurate diagnosis, thereby enhancing the interpretability of the model. The present study develops a vision transformer approach to identify atrial fibrillation based on single-lead ECG data. A residual network (ResNet) approach is also developed for comparison with the vision transformer approach. These models are applied to the Chapman--Shaoxing dataset to classify atrial fibrillation, as well as another common arrhythmia, sinus bradycardia, and normal sinus rhythm heartbeats. The models enable the identification of the key regions of the heartbeat that determine the resulting classification, and highlight the importance of P-waves and T-waves, as well as heartbeat duration and signal amplitude, in distinguishing normal sinus rhythm from atrial fibrillation and sinus bradycardia.  
\end{abstract}



\section{INTRODUCTION}

Atrial fibrillation (AFIB) is one of the most common and clinically significant forms of cardiac arrhythmia. It is characterized by rapid and irregular electrical impulses in the atria, leading to an irregular and often rapid heart rhythm. In the United States, AFIB is estimated to affect 5.2 million individuals, and this number is projected to increase to 12.1 million by 2030. AFIB poses significant risks, including an increased risk of stroke, heart failure, and other cardiovascular complications  \cite{ rakesh_gopinathannair_375a2fab}. The burden on society is estimated to be significant, with AFIB-associated hospitalization and treatment costs amounting to billions of dollars annually \cite{mina_k__chung_5cea7fc4}. 

AI algorithms based on deep learning can effectively analyze and interpret complex ECG waveforms, allowing for the detection and classification of abnormalities such as AFIB. In recent years, researchers have made significant advances in the application of deep neural networks for ECG signal analysis and AFIB detection \cite{zachi_i__attia_6d832af1, akhil_vaid_f8586fad,_zal_y_ld_r_m_f0044257,oliver_faust_592dc342,bambang_tutuko_845bcc82}. With the use of deep neural network models and a carefully-designed network structure, it is possible to achieve classification accuracy comparable to that of human expert cardiologists  \cite{bryan_he_4aa3d1d7}. Moreover, AI algorithms can enable rapid and automated diagnosis, and reduce the risk of human error \cite{joon_myoung_kwon_3b7bf7c6}. 

Researchers have also focused on the interpretability and robustness of neural network models, as their black-box perception can sometimes be a concern for medical professionals \cite{joon_myoung_kwon_3b7bf7c6, mauricio_reyes_877abf86,survey,MOUSAVI2020104057}. By developing explainable AI techniques and incorporating feature extraction methods, it is possible to provide insights into the key features of the data influencing the decision-making process of the neural network \cite{zachi_attia_e73d3d76, Belen}. In addition to deep learning models, classical machine learning approaches employing features extracted from ECG data have also been proposed for ECG classification \cite{jianmin_zheng_361a59f9}. Such models provide insights into the most important features responsible for the resulting diagnosis. However, this approach necessitates feature engineering, which requires prior domain knowledge and expertise for the appropriate selection and extraction of relevant features from the raw signals \cite{joon_myoung_kwon_3b7bf7c6}. There is a need for a visually explainable model that can directly highlight the important regions of the ECG signal, without the prior necessity for extracting features from the ECG signals to serve as model inputs. 

An automated and interpretable method for the identification of arrhythmias from single-lead ECG recordings has the potential to facilitate the remote monitoring of at-risk patients to enable early intervention, aid in understanding the language of the heart, and enhance the physician’s trust to utilize this diagnostic tool. Prior studies have demonstrated that high accuracy is achievable with a single-lead approach \cite{bambang_tutuko_845bcc82}. 

This work develops a vision transformer (ViT) approach for AFIB detection utilizing heartbeat segments contained between 3 consecutive R-peaks in the lead II ECG signals, and demonstrates the ability of the ViT \cite{dosovitskiy2021an} and ResNet \cite{ResNet7780459} models to highlight the key regions of the ECG-recorded heartbeat that characterize AFIB and normal sinus rhythm (SR). In addition to AFIB, another common arrhythmia considered in this study is sinus bradycardia (SB), which is characterized by a slower-than-average heart rhythm. SB has gained attention recently, since bradycardia has been identified as a cardiac manifestation of COVID-19 \cite{douedi2021covid, amaratungacovid}. We used the well-established Chapman--Shaoxing dataset \cite{jianmin_zheng_361a59f9,saira_aziz_98932d60} to develop our explainable multi-class classification approach to distinguish between AFIB, SB, and SR heartbeats.

Our work focuses on lead II with the aim of creating a model that could be transferable to smart watches and other wearable devices for remote monitoring \cite{bambang_tutuko_845bcc82, marc_strik_3b034901}. While previous studies have used fixed segment intervals \cite{PORUMB2020101597} or RR segments \cite{oliver_faust_592dc342} extracted from ECG signals, our study employs RRR segments. This selection was motivated by the fact that an RRR segment contains a complete heartbeat, starting with the P-wave marking the contraction of the heart, and ending with the T-wave associated with the relaxation of the heart. Faust et al. employed RR intervals extracted from 12-lead ECG signals and achieved over 97\% accuracy in classifying AFIB with a ResNet-based model \cite{oliver_faust_592dc342}. While Tutuko et al. demonstrated comparable accuracy of over 96\% in classifying AFIB with a convolutional neural network using lead II alone, they employed longer signals containing 2 or more RR intervals \cite{bambang_tutuko_845bcc82}. Our work achieves comparable accuracy in spite of simultaneously utilizing a single lead and short RRR ECG segments. 

Our work provides an interpretable approach that highlights the key regions of the heartbeat governing the classification, in addition to high classification accuracy. Our results demonstrate that P-waves and T-waves, in addition to heartbeat duration and signal amplitude, play a key role in distinguishing normal SR heartbeats from AFIB or SB arrhythmias.

\section{METHODS}

\subsection{DATASET AND PREPROCESSING}

We utilized the Chapman--Shaoxing database \cite{jianmin_zheng_361a59f9}, and selected patients with AFIB, SB or SR labels. The database contains 10-second ECG recordings taken at a frequency of 500 Hz. We applied the neurokit2 library \cite{Makowski2021neurokit} to the raw ECG data for cleaning and peak detection. The cleaning step performs detrending and denoising via the default 0.5 Hz high-pass 5th order Butterworth filter, followed by powerline filtering. This step is followed by the default R-peak detection. We extracted non-overlapping segments between 3 consecutive R-peaks from the processed ECG data, referred to here as RRR segments, which were then supplied to the classification algorithms. Table \ref{table:dataset} provides the number of patients and RRR segments obtained from the cleaned data for each category.

\begin{table}
\caption{{\small Dataset}}
\vspace{-5pt}
\label{table:dataset}
\begin{center}
{\small
\begin{tabular}{c c c}
\hline
Label & Number of patients & Number of RRR segments\\ [0.5ex]
\hline\hline\\[.05ex]
AFIB & 1654  & 11310 \\ [1ex]
SB & 3765 & 14635 \\ [1ex]
SR & 1789 & 9642 \\ [1ex]
\hline
\end{tabular}}
\end{center}
\end{table}

RRR segments were selected as inputs because each RRR segment contains a complete heartbeat, starting from the P-wave and ending with the T-wave. Moreover, we avoided the use of neighboring RR segments as inputs via our approach of using RRR segments, since adjacent R peaks are likely to be correlated. Each heartbeat extracted from a given patient was associated with the label corresponding to the patient. The application of explainable deep learning models to single heartbeats is intended to elucidate the underlying patterns in the heartbeat associated with different heart conditions. In particular, the use of RRR segments facilitates the visualization of key regions of the individual heartbeats that are associated with the disease diagnosis, leading to an explainable deep learning approach.

\begin{figure}
    \centering
	\includegraphics[width=0.75\columnwidth]{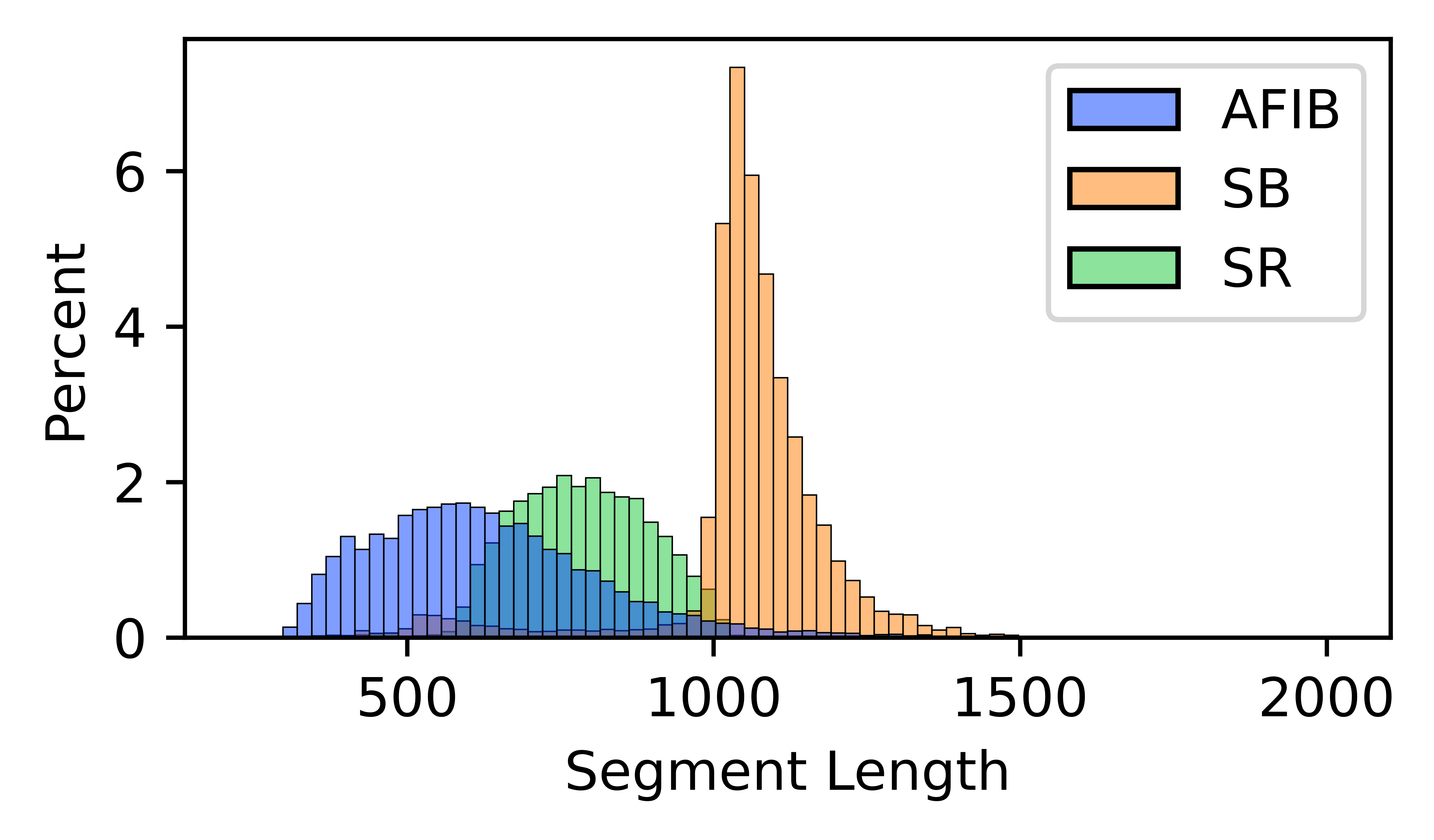}
    \caption{Histogram of RRR segment lengths for AFIB, SB and SR. The $x$-axis represents the segment length, and the $y$-axis represents the percentage in each bin. }\label{fig:histogram_3cl} 
\end{figure}

Since the extracted RRR segments are of varying lengths, the segments were padded with zeros to ensure that the model inputs have the same lengths. Each 10 s ECG recording originally had a length of 5000 points. Figure \ref{fig:histogram_3cl} depicts the histogram of RRR segment lengths for the 3 classes. A maximum length of 1500 was selected, as fewer than 0.1\% of segments exceed this length. 
It is evident from Figure \ref{fig:histogram_3cl} that the 3 classes have distinct distributions of RRR segment lengths, and the segment length may be a factor in classification. 

In prior studies, the ECG signals or extracted segments have been $z$-normalized \cite{PORUMB2020101597} or rescaled by the maximum amplitude \cite{zheng_shaoxingningbo}. However, the ECG amplitude is influenced by the condition of the heart \cite{PRONIEWSKA202345}. In other studies, non-normalized signals were directly employed, even in cases where multiple data sources were used \cite{bambang_tutuko_845bcc82}. In the present study, we utilized both non-normalized and $z$-normalized ECG signals to extract the RRR segments, for comparison.   

\subsection{MODEL DEVELOPMENT}

\begin{figure}
    \centering
	\includegraphics[width=0.6\columnwidth]{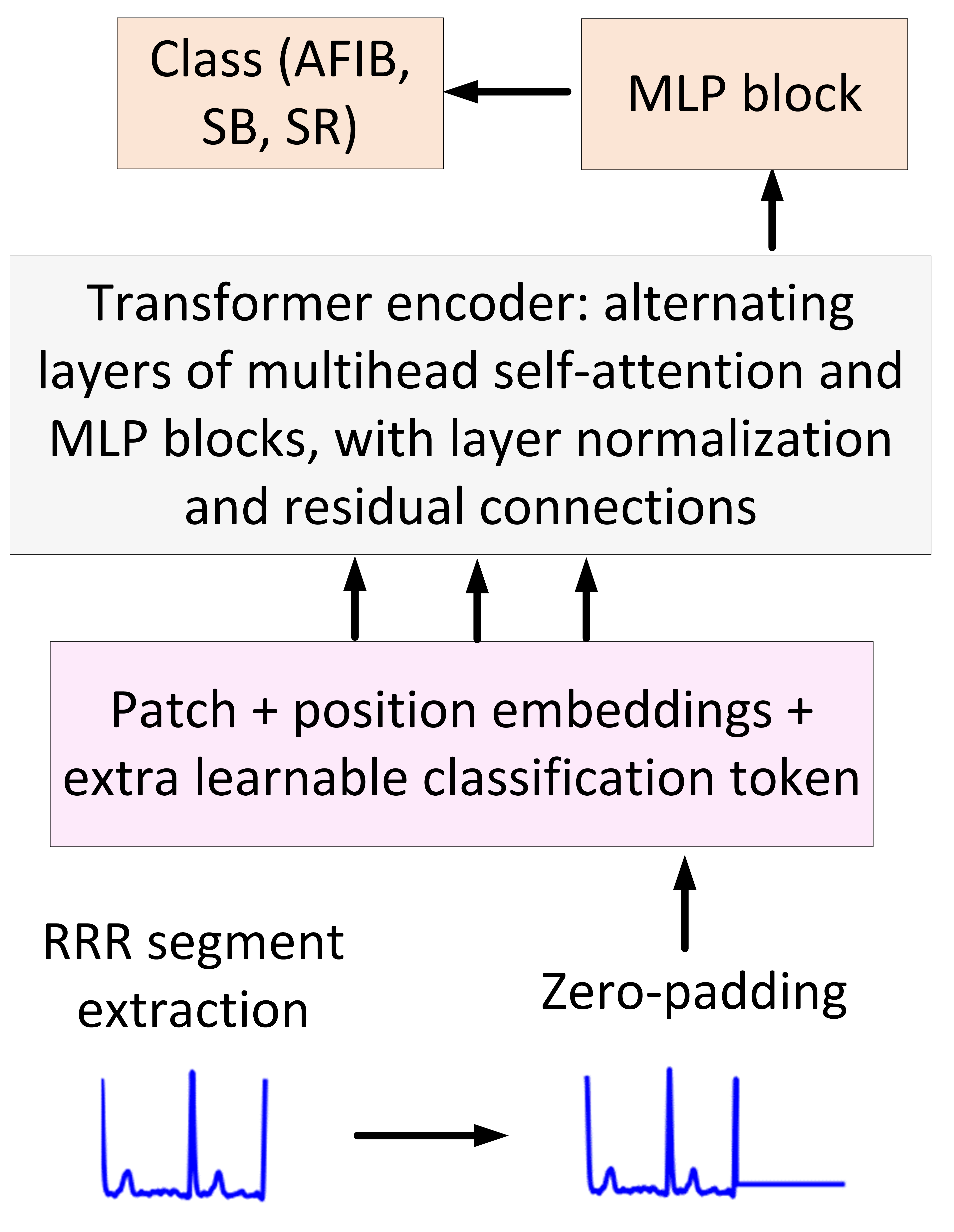}
    \caption{Illustration of the ViT Approach. A detailed diagram of the transformer encoder was presented in \cite{dosovitskiy2021an}.} \label{fig:vitdiag}
\end{figure}

We employed the ViT model including a learnable classification token \cite{dosovitskiy2021an} for the problem of classifying the RRR segments into the 3 classes of labels, AFIB, SB or SR. The model is illustrated in Fig. \ref{fig:vitdiag}. Our ViT approach involved decomposing the RRR segment into patches, and associating learned patch and position embeddings with each patch. The sum of the patch and position embeddings constitutes the input to the transformer encoder, along with the inclusion of an extra learnable classification token. The encoder consists of alternating multihead attention and multilayer perceptron (MLP) blocks, with layer normalization preceding and residual connections following each block. Finally, a layer normalization and MLP block are employed, followed by a dense layer for classification with softmax activation.

The ViT model enables us to visualize the regions of the heartbeat to which the classification token attends via attention heatmaps \cite{NIPS2017_3f5ee243}. For comparison, we employed a ResNet architecture based on ResNet50 \cite{ResNet7780459} and visualized Grad-CAM heatmaps \cite{GradCAM8237336} of the key regions of the signals with regard to the final convolutional layer of the model. The models were developed using the TensorFlow library \cite{tensorflow2015-whitepaper}. Upon comparing the model performance using the signals collected from each single ECG lead, we found that the best-performing leads for the classification model were leads II and aVF, followed by I and V1. For the present study, we selected lead II due to its consistency, reliability, and potential application to wearable devices \cite{marc_strik_3b034901}.

The AFIB, SB and SR heartbeats differ in duration, with AFIB often having a faster-than-average heart rhythm. On the other hand, SB is characterized by a slower-than-average heart rhythm. As described earlier, zero-padding was employed to ensure that the model inputs had the same lengths. However, the signal lengths, reflected in the zero-padded regions, may play a role in the classification results. This was verified in the ViT model, by applying the model with and without the masking of the zeros in the input to the multihead attention layer.

The dataset was split into train, validation and test datasets using a 70-15-15 \%-split. The splitting was performed following the inter-patient paradigm by using group shuffle splitting to ensure that the same patient did not appear in more than one dataset \cite{akhil_vaid_f8586fad,8683140}. Hyperparameter optimization to maximize the validation accuracy was conducted by a combination of manual tuning and the use of the automated keras-tuner library \cite{omalley2019kerastuner}. The hyperparameters tuned in the ViT model were the number of layers, the patch size, the embedding dimension, and the number of units and dropout rate in the MLP layer. The use of 2 or 3 attention heads was explored. In the ResNet model, the ResNet50 architecture was followed by dense and dropout layers preceding the final classification layer with softmax activation, wherein the number of units in the dense layer and the dropout rate were optimized. In addition to our primary ViT and ResNet approaches, we also trained a simpler model for comparison, where a 1D convolutional neural network (CNN) with max pooling followed by a long short-term memory (LSTM) layer was employed, preceding the classification layer with softmax activation. In this CNN--LSTM model, the number of filters and kernel size in the convolutional layer, and the number of units and recurrent dropout rate in the LSTM layer were tuned. The accuracy was reported by averaging the results of 5 independent iterations wherein the model was run after performing train-validation-test splitting. This method is similar to 5-fold cross-validation \cite{PORUMB2020101597}. The model architectures and optimized hyperparameters are summarized in Table \ref{table:model_arch}. All computations were performed on Amazon Web Services (AWS) GPU instances.

\begin{table}
\caption{{\small Model Architectures}}
\label{table:model_arch}
\begin{center}
{\small
\begin{tabular}{p{0.1\columnwidth} p{0.8\columnwidth}}
\hline
Model & Hyperparameters\\ [0.5ex]
\hline\hline\\[.05ex]
ResNet & ResNet50 architecture followed by dense layer with 64 units and dropout rate of 0.2\\ [1ex]
ViT & 2 or 3 attention layers, 2 or 3 attention heads, patch size of 30, embedding dimension of 16 and MLP layer with 128 units\\ [1ex]
CNN--LSTM & CNN layer with 32 filters, kernel size of 3, max pooling with pool size of 2, followed by 96 LSTM units with recurrent dropout rate of 0.2\\ [1ex]
\hline
\end{tabular}}
\end{center}
\end{table}

\section{RESULTS}

The overall accuracies on the test dataset, averaged over 5 independent iterations of data splitting and model fitting, are reported in Table \ref{table:model_perf}. The ViT model based on the non-normalized dataset employed 3 attention layers, whereas 2 attention layers were found to be sufficient to optimize the model performance for the $z$-normalized dataset. The accuracies and other metrics for the ViT models are based on utilizing 2 attention heads. The accuracy upon employing 3 attention heads did not show notable improvement, at 0.9255 for the unmasked ViT model using non-normalized data. 

\begin{table}
\caption{{\small Overall model accuracy for RRR segment classification}}
\label{table:model_perf}
\begin{center}
{\small 
\begin{tabular}{c c c}
\hline
Model & Normalization & Accuracy\\ [0.5ex]
\hline\hline\\[.05ex]
& Non-normalized & 0.9613 \\[.5ex]
\raisebox{.75ex}{ResNet} & z-normalized & 0.8755 \\[1ex]
& Non-normalized & 0.9246 \\[.5ex]
\raisebox{.75ex}{ViT, unmasked} & z-normalized & 0.8550 \\[1ex]
& Non-normalized & 0.9073 \\[.5ex]
\raisebox{.75ex}{ViT, masked} & z-normalized & 0.8918 \\[1ex]
& Non-normalized & 0.8878 \\[.5ex]
\raisebox{.75ex}{CNN--LSTM} & z-normalized & 0.8285 \\[1ex]
\hline
\end{tabular}}
\end{center}
\end{table}

Table \ref{table:model_perf} demonstrates that the performance of the ViT model declined upon masking the zeros in the input segments. This indicates that the segment length is an important feature, and is captured indirectly in the model by attending to the zeros in the input data. Similarly, the model performance declined upon employing $z$-normalized ECG data, indicating that the signal amplitudes are an important indicator of the heart condition. The CNN--LSTM model had the lowest accuracy and longest run time among the tested approaches, and was not pursued further.

Table \ref{table:model_rrr_perf_labels} presents several performance metrics by label for the ResNet model and the unmasked ViT model using non-normalized data, based on a one-versus-rest calculation. The metrics are obtained at a patient-level by applying a majority vote of RRR heartbeat segment labels for each patient. The resulting patient-level metrics are listed in Table \ref{table:model_patient_perf_labels}. The gap between the heartbeat-level metrics and patient-level metrics is not very large, suggesting that a given patient's heart condition is reflected in many of the heartbeats extracted from that patient's ECG recording. This observation provides justification for performing classification based on individual heartbeats.

\begin{table}
\caption{{\footnotesize Model performance by label for RRR segment classification}}
\label{table:model_rrr_perf_labels}
{\tiny
\begin{tabularx}{\columnwidth}{c c c c c c c c}
\hline
Model & Label & Accuracy & Specificity &  Sensitivity & Precision & F1-score & AUC\\[0.5ex]
\hline\hline\\[.05ex]
& AFIB & 0.9827 & 0.9894 & 0.9695 & 0.9789 & 0.9741 & 0.9978\\[.5ex]
ResNet & SB & 0.9722 & 0.9742 & 0.9691 & 0.9614 & 0.9652 & 0.9957\\[.5ex]
& SR & 0.9678 & 0.9781 & 0.9392 & 0.9400 & 0.9394 & 0.9922\\[1.5ex]
& AFIB & 0.9460 & 0.9591 & 0.9203 & 0.9201 &  0.9197 & 0.9884\\[.5ex]
ViT (unmasked) & SB & 0.9622 & 0.9651 & 0.9578 & 0.9480 & 0.9528 & 0.9919\\[.5ex]
& SR & 0.9409 & 0.9629 & 0.8802 & 0.8964 & 0.8878 & 0.9807\\[1.5ex]
\hline
\end{tabularx}}
\end{table}

\begin{table}
\caption{{\small Model performance by label for patient classification}}
\label{table:model_patient_perf_labels}
\tiny{
\begin{tabularx}{\columnwidth}{c c c c c c c c}
\hline
Model & Label & Accuracy & Specificity &  Sensitivity & Precision & F1-score & AUC\\[0.5ex]
\hline\hline\\[.05ex]
& AFIB & 0.9928 & 0.9922 & 0.9947 & 0.9763 & 0.9853 & 0.9998\\[.5ex]
ResNet & SB & 0.9810 & 0.9691 & 0.9924 & 0.9710 & 0.9816 & 0.9990\\[.5ex]
& SR & 0.9811 & 0.9980 & 0.9296 & 0.9936 & 0.9604 & 0.9981\\[1.5ex]
& AFIB & 0.9678 & 0.9705 & 0.9597 & 0.9132 &  0.9355 & 0.9953\\[.5ex]
ViT (unmasked) & SB & 0.9754 & 0.9642 & 0.9862 & 0.9664 & 0.9762 & 0.9983\\[.5ex]
& SR & 0.9636 & 0.9912 & 0.8794 & 0.9705 & 0.9225 & 0.9946\\[1.5ex]
\hline
\end{tabularx}}
\end{table}

The ResNet model is able to achieve overall and one-versus-rest classification accuracies of over 96\%, whereas the unmasked ViT model lags behind, with an overall accuracy of around 92-93\%, as seen from Table \ref{table:model_perf} and the preceding discussion. This is expected, since the ViT model was proposed for large dataset sizes exceeding 10s of millions \cite{dosovitskiy2021an}. However, the ViT model is expected to perform better as larger datasets become available \cite{dosovitskiy2021an}, and also provides a clear illustration of the key regions of the heartbeat influencing the resulting classification via the attention heatmaps obtained from each attention head. 

The attention maps were obtained from the ViT self-attention calculation by employing the computed attention of the classification token toward the rest of the sequence, averaged for the corresponding labels, and scaled to lie between 0 and 1. Since the RRR segments and corresponding attention maps are of variable lengths, the attention heatmaps, as well as the RRR segments, were extrapolated to the maximum segment length of 1500, and averaged to produce average attention maps and RRR heartbeats. A similar procedure was applied to plot the Grad-CAM heatmaps.

\begin{figure}
	\subfloat[][$1^{\text{st}}$ attention head, correctly classified AFIB]{\includegraphics[width=0.49\columnwidth]{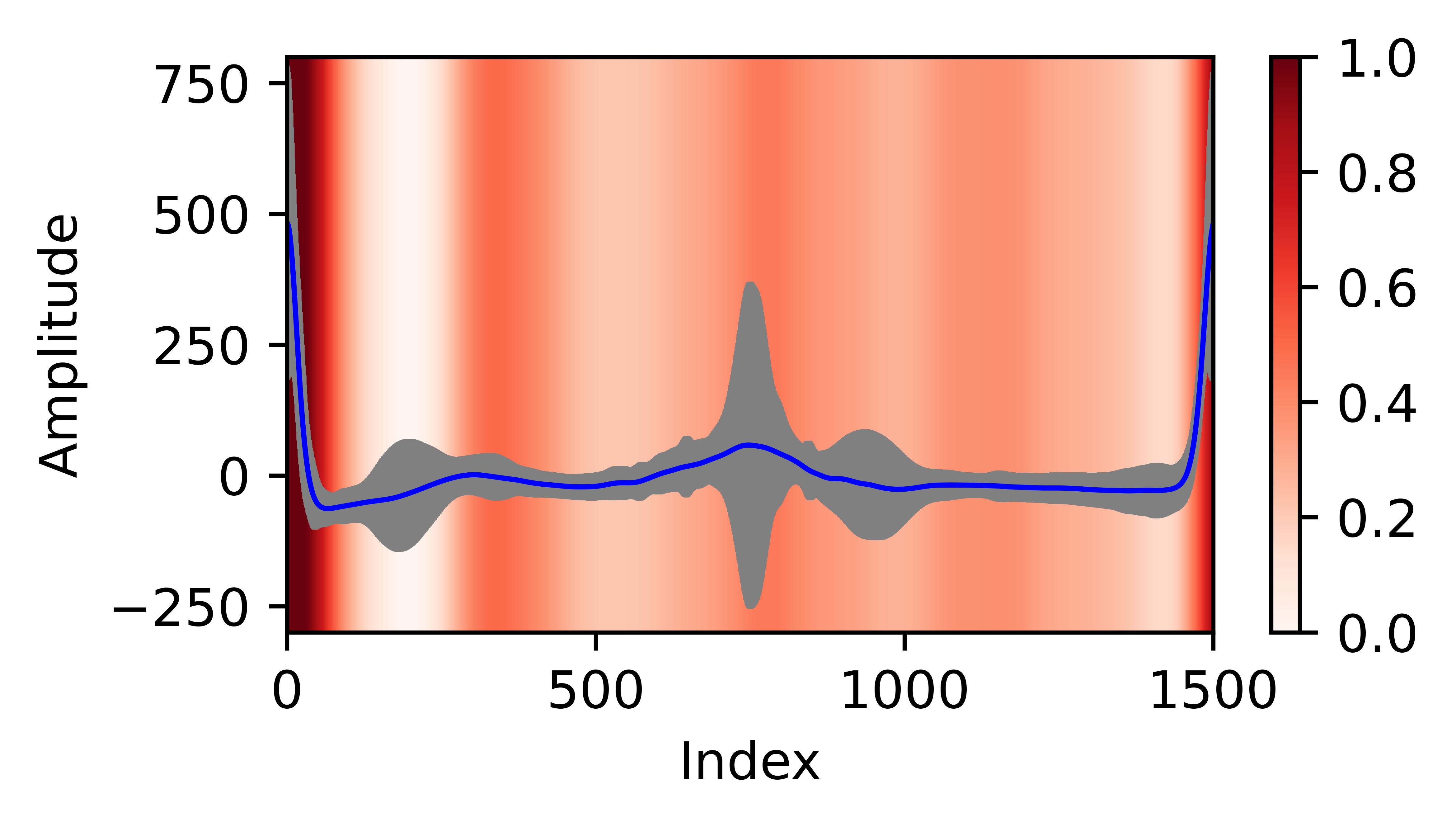}}\label{fig:vit_2attheads_correct_AFIB_head0}
	\subfloat[][$2^{\text{nd}}$ attention head, correctly classified AFIB]{\includegraphics[width=0.49\columnwidth]{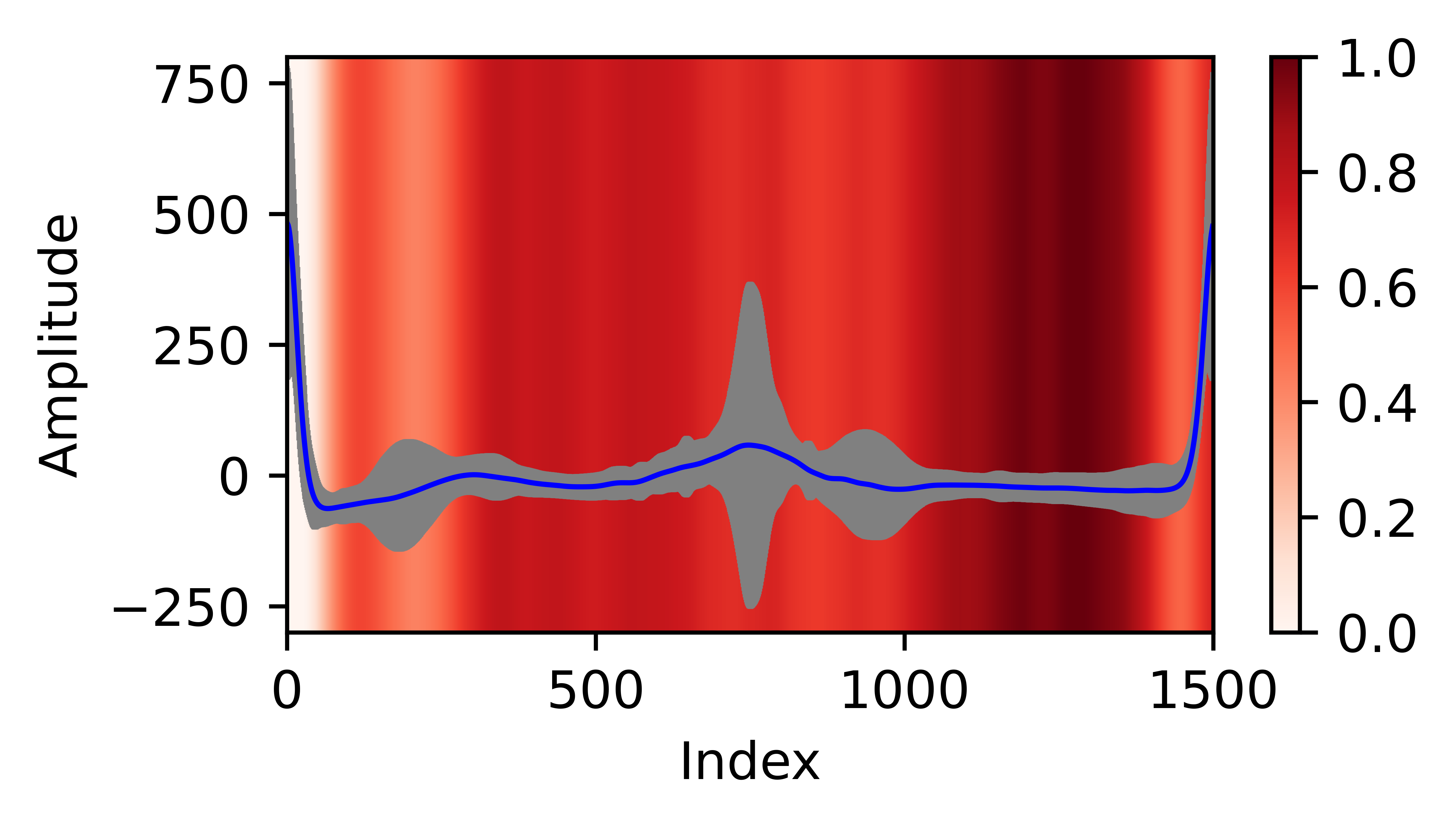}}\label{fig:vit_2attheads_correct_AFIB_head1} \\
    \caption{{\small Attention heatmaps from the ViT model using 2 attention heads, based on correctly classified AFIB cases. The blue line represents the average signal amplitude, with the gray region corresponding to $\pm$ 1 standard deviation. The $y$-axis represents the amplitude in $\mu$V, while the $x$-axis represents the index.}}
    \label{fig:vit_2heads_corr_AFIB}
\end{figure}

\begin{figure}
	\subfloat[][$1^{\text{st}}$ attention head, correctly classified SB]{\includegraphics[width=0.49\columnwidth]{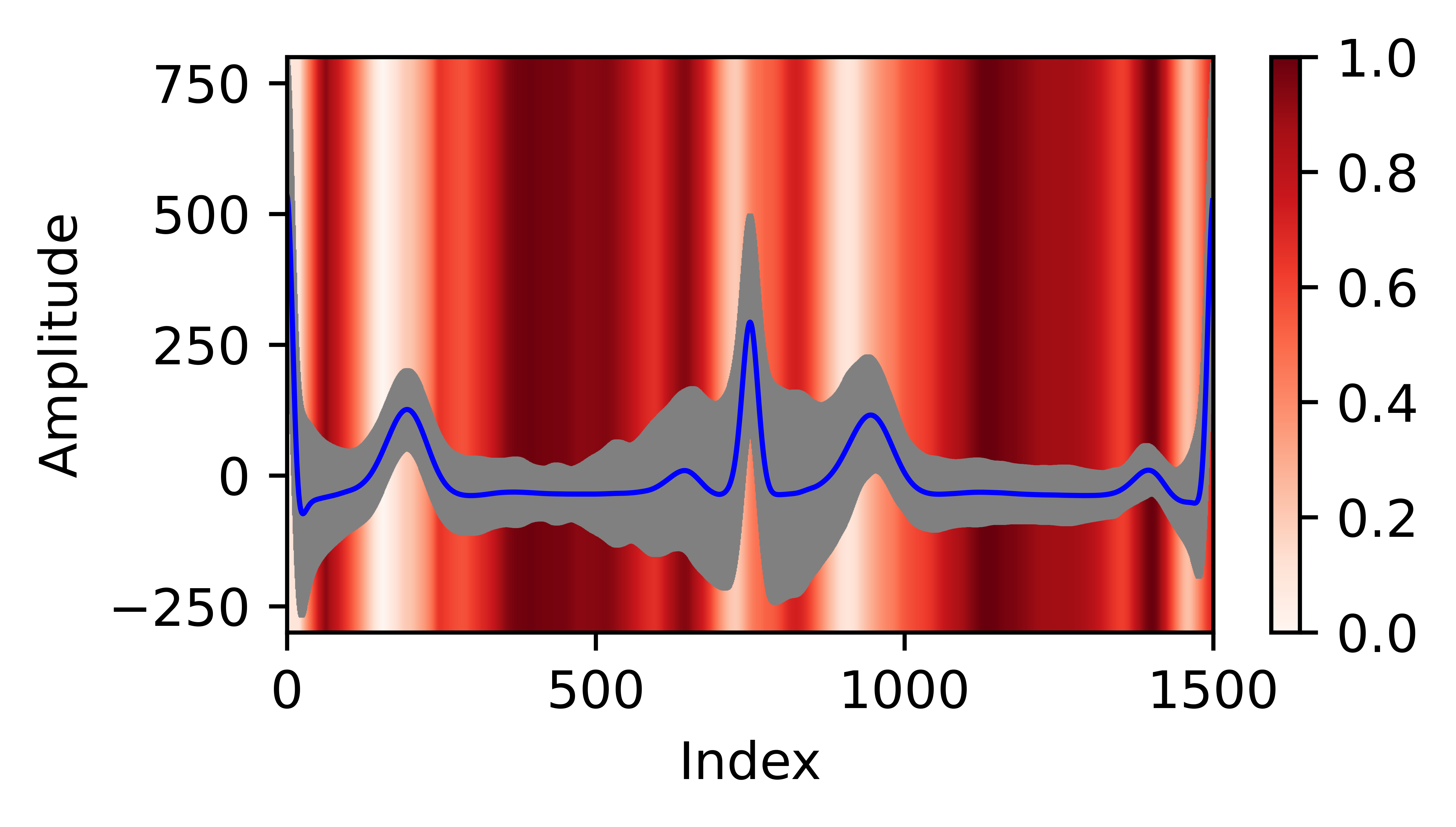}}\label{fig:vit_2attheads_correct_SB_head0}
	\subfloat[][$2^{\text{nd}}$ attention head, correctly classified SB]{\includegraphics[width=0.49\columnwidth]{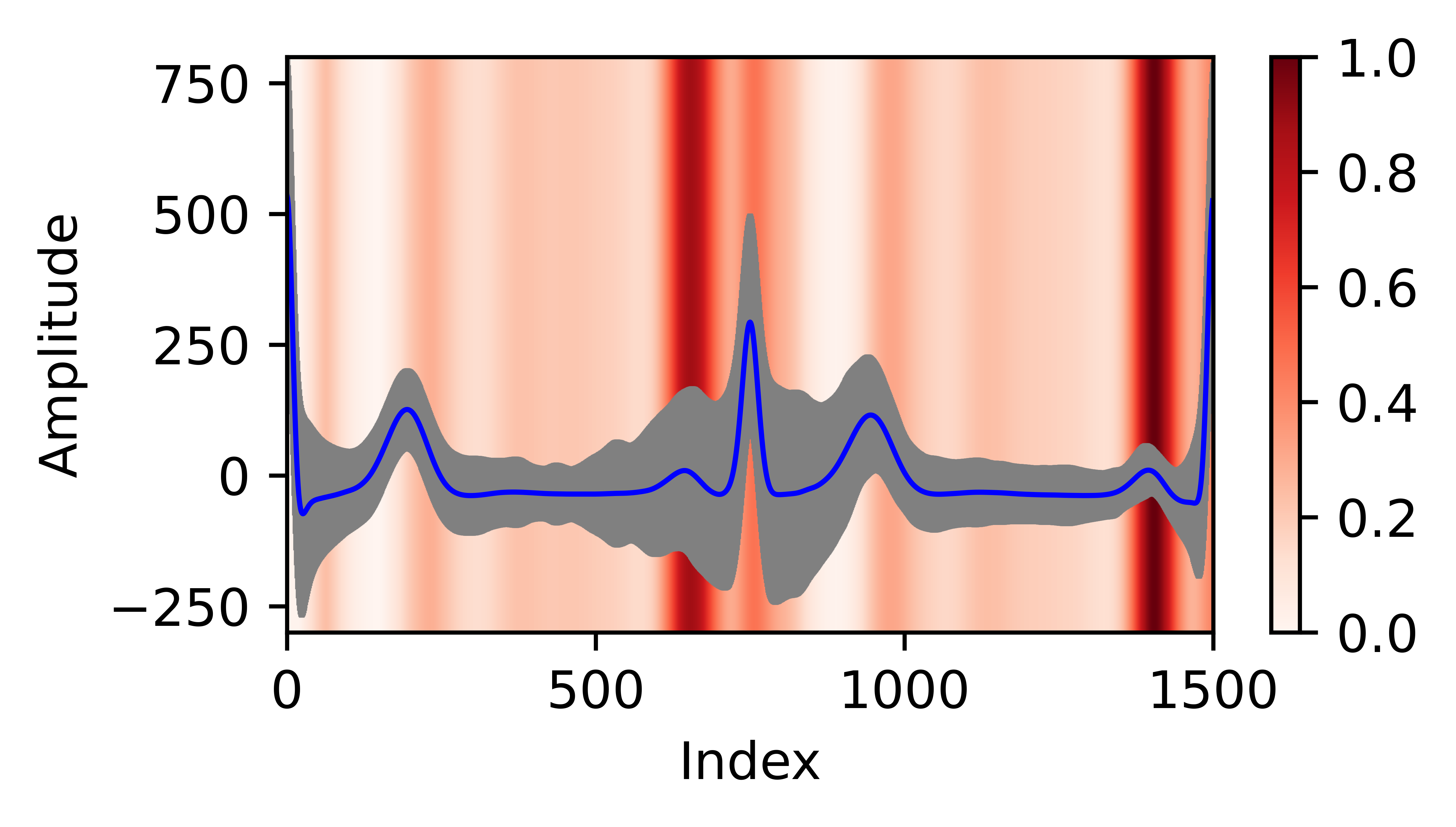}}\label{fig:vit_2attheads_correct_SB_head1}
    \caption{{\small Attention heatmaps from the ViT model using 2 attention heads, based on correctly classified SB cases. The blue line represents the average signal amplitude, with the gray region corresponding to $\pm$ 1 standard deviation. The $y$-axis represents the amplitude in $\mu$V, while the $x$-axis represents the index.}}
    \label{fig:vit_2heads_corr_SB}
\end{figure}

\begin{figure}
	\subfloat[][$1^{\text{st}}$ attention head, correctly classified SR]{\includegraphics[width=0.49\columnwidth]{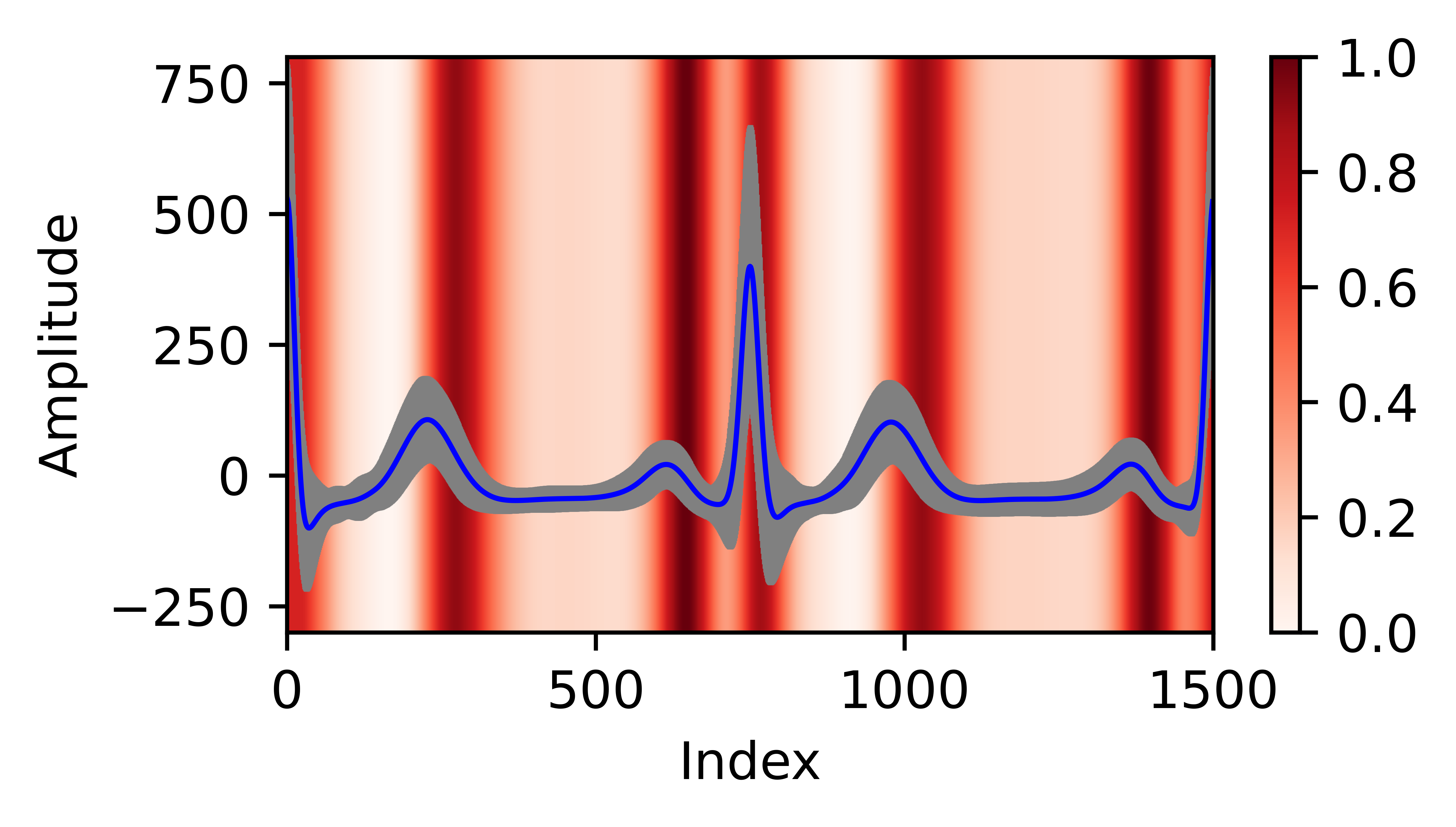}}\label{fig:vit_2attheads_correct_SR_head0}
	\subfloat[][$2^{\text{nd}}$ attention head, correctly classified SR]{\includegraphics[width=0.49\columnwidth]{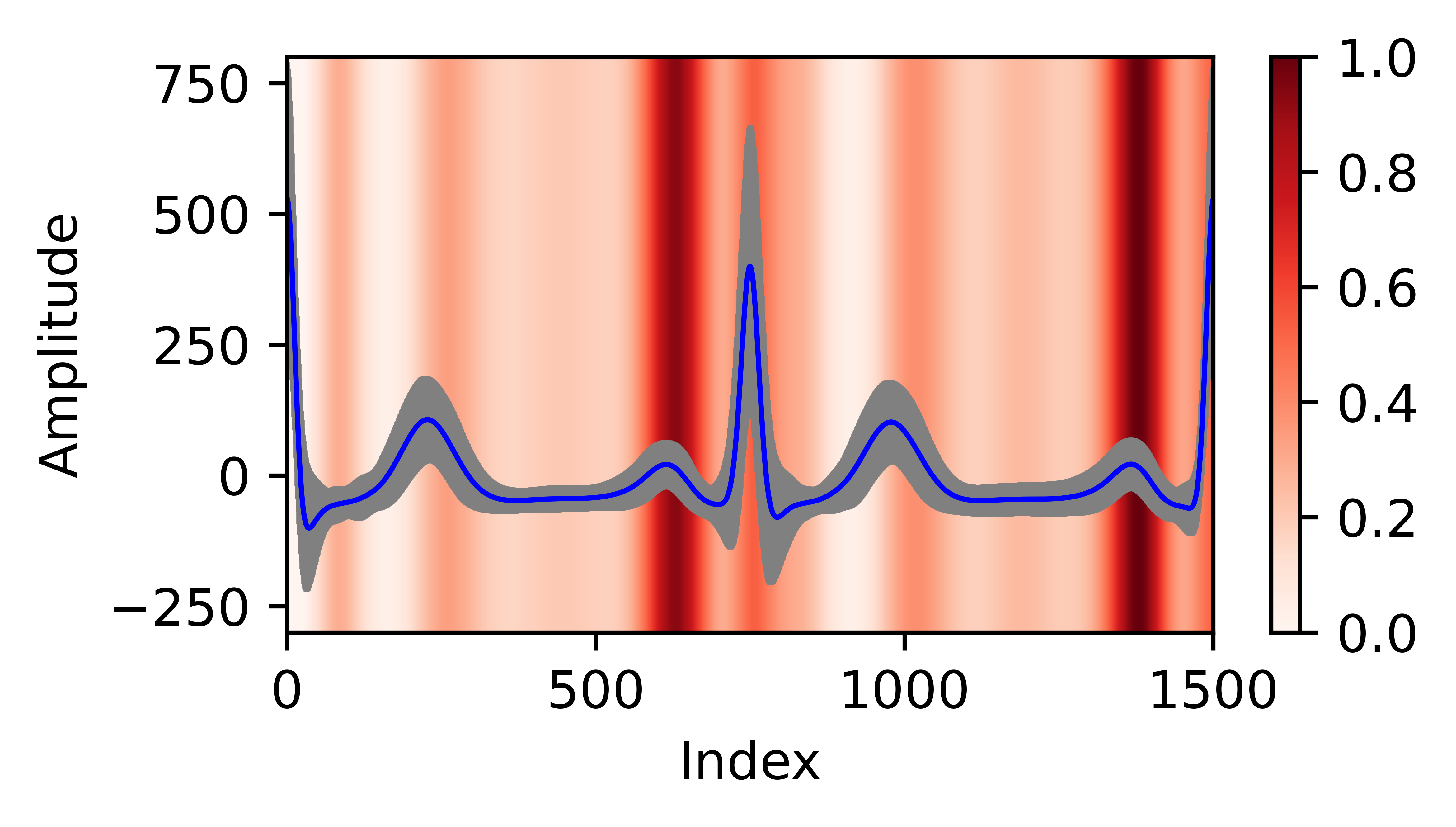}}\label{fig:vit_2attheads_correct_SR_head1}
    \caption{{\small Attention heatmaps from the ViT model using 2 attention heads, based on correctly classified SR cases. The blue line represents the average signal amplitude, with the gray region corresponding to $\pm$ 1 standard deviation. The $y$-axis represents the amplitude in $\mu$V, while the $x$-axis represents the index.}}
    \label{fig:vit_2heads_corr_SR}
\end{figure}

Figures \ref{fig:vit_2heads_corr_AFIB}, \ref{fig:vit_2heads_corr_SB} and \ref{fig:vit_2heads_corr_SR} illustrate the averaged, scaled attention heatmaps from the unmasked ViT model using non-normalized input data for correctly classified AFIB, SB and SR cases, respectively, superimposed upon the corresponding average RRR segments. These attention heatmaps demonstrate that the model attends primarily to the P-waves, followed by the T-waves, and finally the R-peaks, for SR identification. This can be understood by the absence of P-waves in most AFIB cases. In addition, T-wave inversion has been reported to occur concurrently with AFIB in patients with long-standing hypertension \cite{Hong_AFib_TWI}. Figure \ref{fig:vit_2heads_corr_AFIB} depicts a large variation in the T-waves in AFIB cases, consistent with the inversion of the T-waves in several patients. The irregularity in the QRS complex is also evident, as is the absence of P-waves in AFIB cases. In the SB cases, the model attends to the region of the P-waves, as well as preceding the P-waves. SB is characterized by a longer heartbeat, and the segment length plays a role in the classification model. In addition, the P-wave amplitudes at the lower heart rates associated with SB cases are generally lower relative to normal SR, whereby the amplitude also plays a role in the classification result. In the AFIB cases, the model attends to the region where a P-wave would have otherwise been expected, but is generally absent in the case of AFIB, as well as the location of the R-peaks, which is indicative of the irregularity in the QRS complex in AFIB.  

\begin{figure}
    \centering
	\subfloat[][Correctly classified AFIB]{\includegraphics[width=0.49\columnwidth]{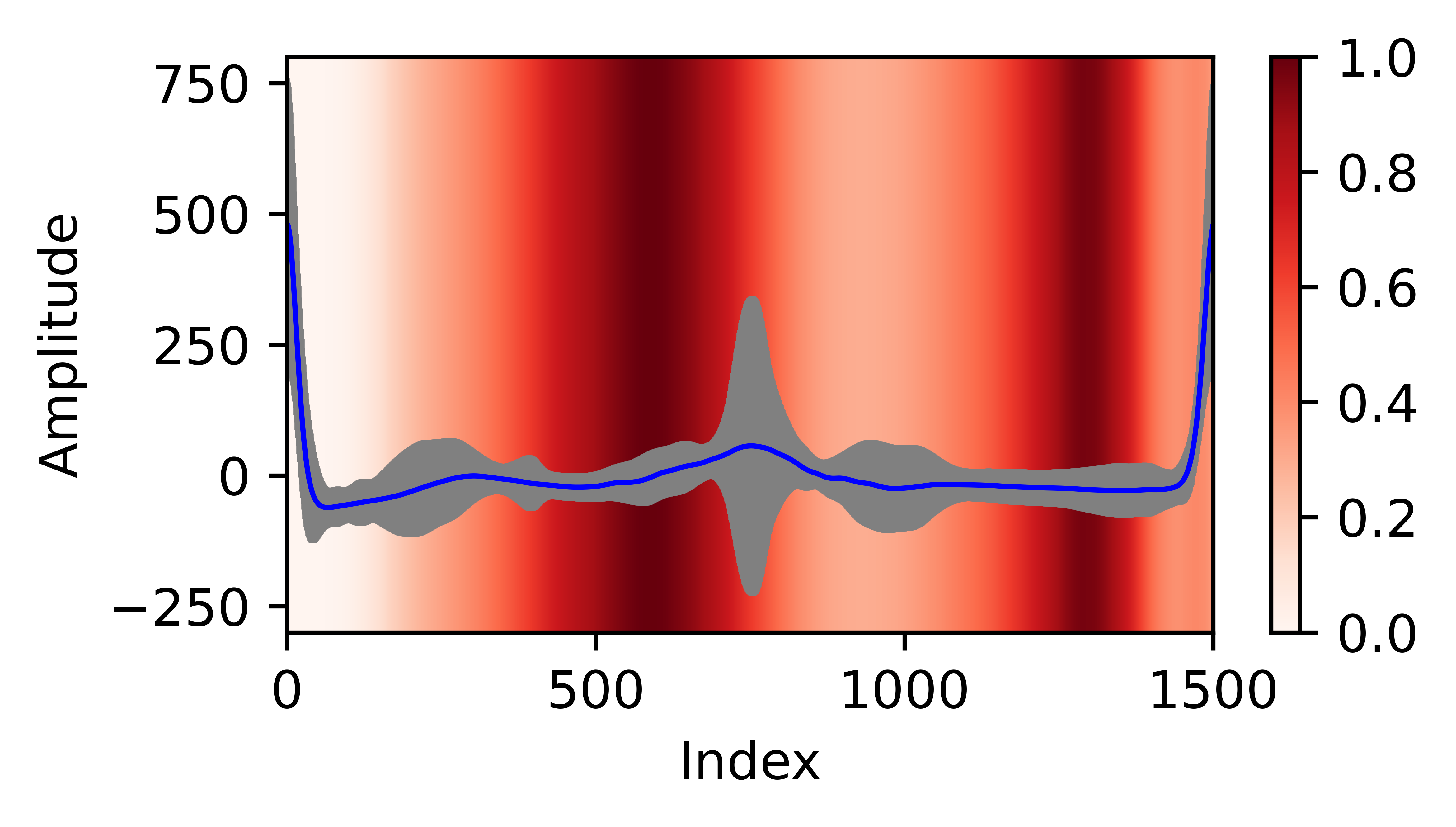}}\label{fig:gradcam_correct_AFIB}
	\subfloat[][Correctly classified SB]{\includegraphics[width=0.49\columnwidth]{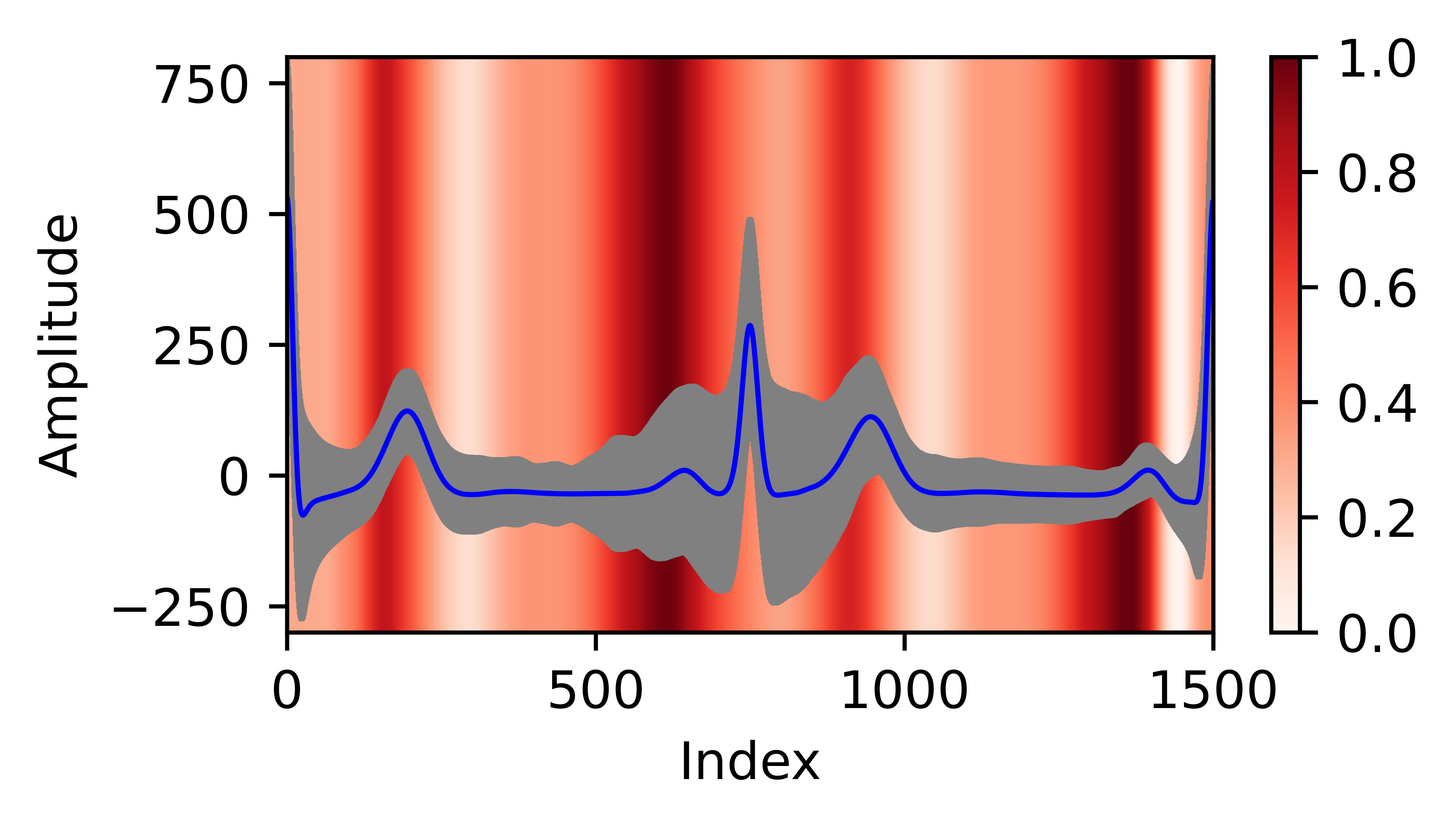}}\label{fig:gradcam_correct_SB}
	\subfloat[][Correctly classified SR]{\includegraphics[width=0.49\columnwidth]{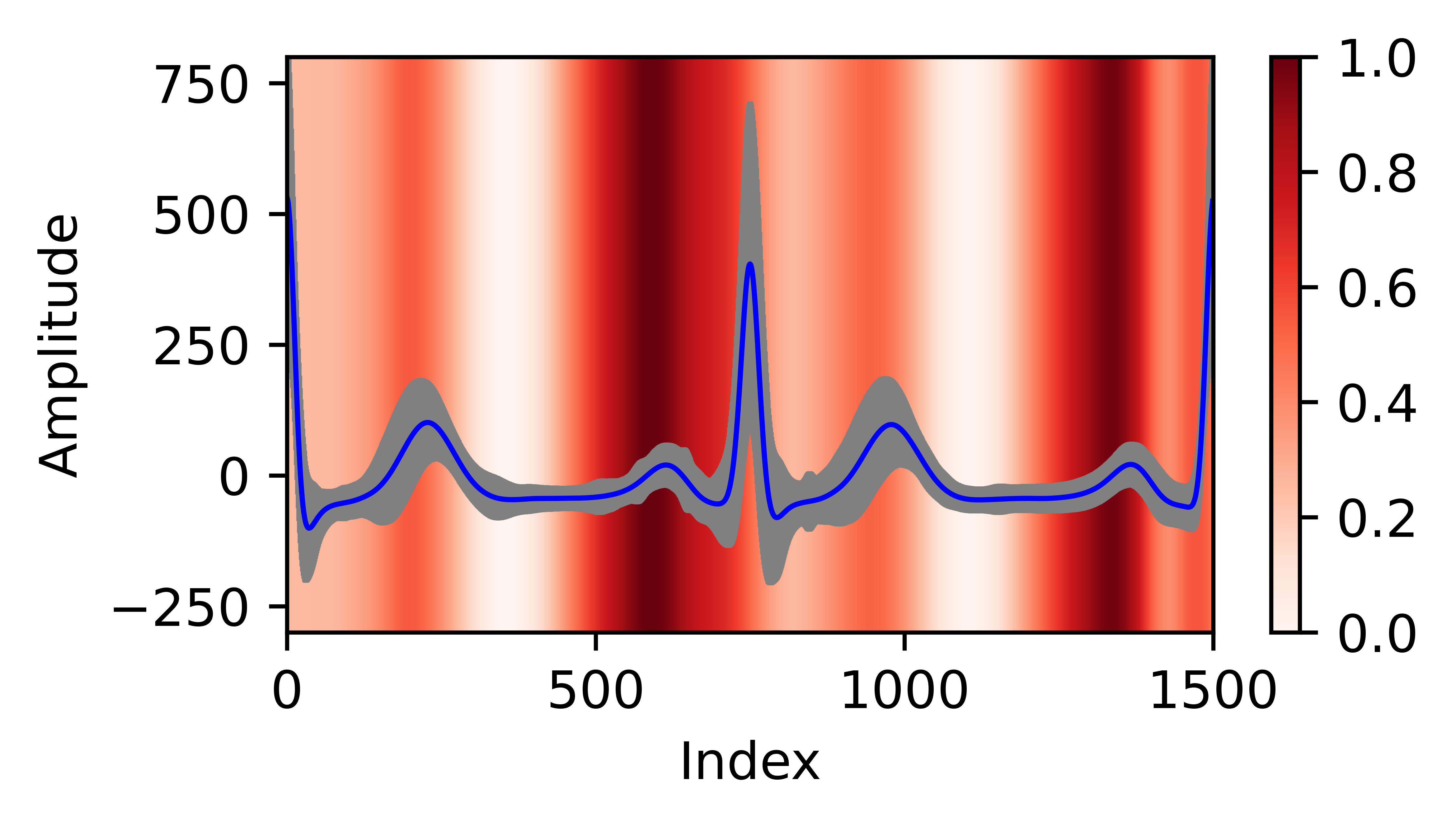}}\label{fig:gradcam_correct_SR} 
    \caption{{\small Grad-CAM heatmaps from the ResNet model, based on correctly classified AFIB, SB and SR cases. The blue line represents the average signal amplitude, with the gray region corresponding to $\pm$ 1 standard deviation. The $y$-axis represents the amplitude in $\mu$V, while the $x$-axis represents the index.}}
    \label{fig:gradcam_correct}
\end{figure}

Figure \ref{fig:gradcam_correct} depicts the averaged and scaled Grad-CAM heatmaps from the ResNet model for correctly classified AFIB, SB and SR cases, respectively. Similar to the attention maps, the Grad-CAM maps emphasize the P-waves and T-waves for the SB and SR cases, and the region where P-waves would have been expected for the AFIB cases. The results obtained from the ViT attention maps and ResNet Grad-CAM maps are consistent with each other. However, the ViT model is computationally much faster. For example, one iteration of model fitting and inference, followed by the attention or Grad-CAM calculation on the test data, took 142 seconds for the ViT model and 1126 seconds for the ResNet model on an AWS ml.g5.24xlarge NVIDIA GPU instance. The CNN--LSTM model had a much longer run time of 14561 seconds for one iteration of model fitting and inference on the same instance.  The ViT model is more readily parallelizable due to its multihead attention architecture, which provides a computational benefit relative to ResNet or sequential deep learning architectures, especially for larger datasets \cite{dosovitskiy2021an}. On account of this computational advantage, the ViT approach has the potential to provide accelerated inference, either on the cloud, or real-time on edge devices \cite{Li10181174} for applications to wearable smart devices.

We also examined the attention heatmaps and Grad-CAM maps for the misclassified cases (not shown here). The misclassified cases appear much noisier, which may have resulted in the incorrect identification of spurious P-waves and T-waves.


These results demonstrate the potential for automated classification of heart conditions by deep learning models, based on the identification of waveforms in the heartbeats, in addition to the heartbeat lengths and amplitudes. Furthermore, the ViT and ResNet approaches enable the identification of the key regions of the heartbeat responsible for the classification result, thereby rendering the results more easily explainable.

\section{CONCLUSIONS}
This paper aims at developing a reliable and interpretable approach for AFIB detection using short single-lead ECG signals. For this purpose, we developed ViT and ResNet models for the classification of AFIB, SB and SR cases using RRR-segmented lead II ECG signals from the Chapman--Shaoxing dataset. The ViT model was evaluated versus the ResNet model. The ResNet model achieved an overall accuracy of over 96\%, whereas the ViT model provided an accuracy in the range 92--93\%. This is expected, since the ViT model was developed for much larger datasets in the size range of 10s of millions \cite{dosovitskiy2021an}. We are currently addressing this issue by identifying and analyzing larger datasets to improve the performance of our ViT approach.

The attention maps and Grad-CAM maps derived from the ViT and ResNet models illustrate the regions of the heartbeats that govern the resulting classification. The heatmaps emphasize the role played by P-waves and T-waves, in addition to other factors including the segment lengths and amplitudes, in distinguishing between AFIB, SB and SR cases. 

The explainable deep learning models explored in this work facilitate the detection of atrial fibrillation, as well as other arrhythmias, from single-lead ECG data, while highlighting the regions of the heartbeat that determine the diagnosis. These models can potentially be employed in conjunction with wearable ECG devices for remote monitoring of patients to facilitate early detection and intervention. Further work will include the use of larger datasets, particularly for the ViT model, which relies on the availability of large amounts of data. In addition, the application of explainable deep learning models to other heart conditions will be explored in the future.

\addtolength{\textheight}{-1cm}   





\def\baselinestretch{0.91}
\bibliographystyle{IEEEtran}

\bibliography{refs_short}%

\end{document}